\documentclass[twocolumn,amsmath,amssymb,aps,preprintnumbers,prl]{revtex4-1}
\usepackage{times}
\usepackage{amsmath}
\usepackage{amsfonts}
\usepackage{graphicx}
\usepackage{dcolumn}
\usepackage{bm}
\usepackage{color}

\def\beq{\begin{equation}}
\def\eeq{\end{equation}}
\def\bea{\begin{eqnarray}}
\def\eea{\end{eqnarray}}
\def\beqn{\begin{eqnarray}} \def\eeqn{\end{eqnarray}}
\def\beeq{\begin{eqnarray}}
\def\eeeq{\end{eqnarray}}

\def\nn{\nonumber}
\def\Eq#1{Eq.~(\ref{#1})}

\def\td#1{\tilde{\delta}\left(#1\right)}

\def\qb{\mathbf{q}}

\def\ii{\imath 0}

\begin{document}

\preprint{IFIC/19-22}

\title{Causality, unitarity thresholds, anomalous thresholds and infrared singularities from the loop-tree duality at higher orders}

\author{J. Jes\'us Aguilera-Verdugo~$^{(a)}$} \email{jesus.aguilera@ific.uv.es}
\author{F\'elix Driencourt-Mangin~$^{(a)}$} \email{felix.dm@ific.uv.es} 
\author{Judith Plenter~$^{(a)}$} \email{judith.plenter@ific.uv.es}
\author{Selomit Ram\'{\i}rez-Uribe~$^{(a)}$} \email{norma.selomit.ramirez@ific.uv.es}
\author{Germ\'an Rodrigo~$^{(a)}$} \email{german.rodrigo@csic.es}
\author{Germ\'an F. R. Sborlini~$^{(a)}$} \email{german.sborlini@ific.uv.es}
\author{William J. Torres Bobadilla~$^{(a)}$} \email{william.torres@ific.uv.es}
\author{Szymon Tracz~$^{(a)}$} \email{szymon.tracz@ific.uv.es}
\affiliation{ ${}^{a}$ Instituto de F\'{\i}sica Corpuscular, Universitat de Val\`{e}ncia -- Consejo Superior de Investigaciones Cient\'{\i}ficas, 
Parc Cient\'{\i}fic, E-46980 Paterna, Valencia, Spain. }

\date{\today}

\begin{abstract}
We present the first comprehensive analysis of the unitarity thresholds and anomalous thresholds 
of scattering amplitudes at two loops and beyond based on the loop-tree duality, and show how non-causal 
unphysical thresholds are locally cancelled in an efficient way when the forest of all the dual on-shell cuts is considered as one. 
We also prove that soft and collinear singularities at two loops and beyond are restricted to a compact region of the loop three-momenta, 
which is a necessary condition for implementing a local cancellation of loop infrared singularities with the ones appearing in real emission; without relying on a subtraction formalism. 
\end{abstract}

\maketitle


\section{Introduction}

It is common knowledge that scattering amplitudes in Quantum Field Theory can be reconstructed from their singularities. 
While tree-level amplitudes only have poles loop amplitudes develop branch cuts, corresponding to 
discontinuities associated with physical thresholds as well as infrared (IR) singularities in the soft and collinear limits.
The singular IR behavior of QCD amplitudes is well-known through general factorization formulas~\cite{Catani:1998bh,Becher:2009cu}.
The physical consequences of the emergence of unitarity thresholds~\cite{Cutkosky:1960sp, Mandelstam:1960zz}, anomalous thresholds and more generally Landau singularities~\cite{Landau:1959fi,Mandelstam:1960zz,Cutkosky:1961,Coleman:1965xm,Kershaw:1971rc,Deshpande:1991pn,Frink:1997sg,Goria:2008ny,Dennen:2016mdk,
Chin:2018puw,Passarino:2018wix,Gomez:2019sxl},
for specific kinematical configurations, have also been extensively discussed in the literature. Indeed
a thorough knowledge of the singular structure of scattering amplitudes is a prerequisite 
for obtaining theoretical predictions of physical observables.

The Loop-Tree Duality (LTD)~\cite{Catani:2008xa,Bierenbaum:2010cy,Bierenbaum:2012th,Buchta:2014dfa,Buchta:2015wna,Driencourt-Mangin:2017gop,Driencourt-Mangin:2019aix} is a powerful framework to analyze the 
singular structure of scattering amplitudes directly in the loop momentum space. 
The LTD representation of a one-loop scattering amplitude is given by
\bea
{\cal A}^{(1)} (\{p_n\}_N) &=& - \int_{\ell} {\cal N}(\ell,\{p_n\}_N) \otimes G_D(\alpha)~, \nn \\
G_D(\alpha) &=& \sum_{i\in \alpha} \td{q_i} \prod_{j\ne i}G_D(q_i; q_j)~,
\label{eq:LTDoneloop}
\eea
where ${\cal N} (\ell,\{p_n\}_N)$ is the numerator of the integrand that depends on the loop momentum $\ell$ and 
the $N$ external momenta $\{p_n\}_N$ (for explanations on the notation used consult \cite{Catani:2008xa,Driencourt-Mangin:2019aix}). The delta function $\td{q_i} = \imath 2 \pi \, \theta(q_{i,0}) \, \delta (q_i^2-m_i^2)$
sets on-shell the internal propagator with momentum $q_i=\ell+k_i$ and selects its positive energy mode, $q_{i,0}>0$.
At one-loop, $\alpha=\{1, \cdots, N\}$ labels all the internal momenta, and \Eq{eq:LTDoneloop} is the sum of $N$ single-cut
dual amplitudes. The dual propagators,
\beq
G_D(q_i;q_j) = \frac{1}{q_j^2-m_j^2 - \ii \, \eta \cdot k_{ji}}~,  
\eeq
differ from the usual Feynman propagators only by the imaginary prescription that now depends on 
$\eta\cdot k_{ji}$, with $k_{ji} = q_j-q_i$. Notice that the dual propagators are implicitly linear in the loop momentum due to the on-shell conditions. Though the vector $\eta$ is mostly arbitrary -- it only has to be future-like -- 
the most convenient choice is $\eta = (1, {\bf 0})$. This election is equivalent to integrating out the energy 
component of the loop momentum which renders the remaining integration domain Euclidean.  

\begin{figure}[th]
\begin{center}
\includegraphics[width=0.20\textwidth]{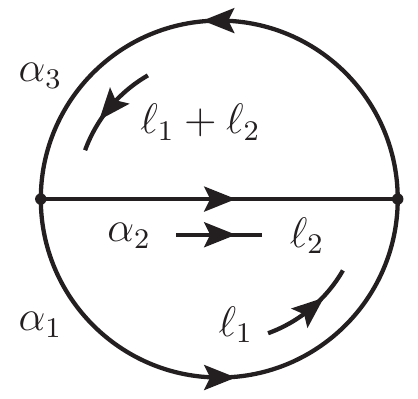}
\caption{Momentum flow of a two-loop Feynman diagram.
\label{fig:twoloop}}
\end{center}
\end{figure}

The master dual representation of a two-loop scattering amplitude is~\cite{Bierenbaum:2012th,Driencourt-Mangin:2019aix}
\bea
&& {\cal A}^{(2)}(\{p_n\}_N) = \int_{\ell_1}\int_{\ell_2} {\cal N}(\ell_1,\ell_2,\{p_n\}_N) \otimes 
\nn \\ && \qquad [ G_D(\alpha_1) \, G_D(\alpha_2\cup \alpha_3) 
+ G_D(-\alpha_2\cup \alpha_1) G_D(\alpha_3) \nn \\ 
&& \qquad - G_D(\alpha_1) \,G_F(\alpha_2) \, G_D(\alpha_3) ]~,
\label{eq:LTDtwoloop}
\eea
where the internal momenta $q_i = \ell_1+k_i$, $q_j = \ell_2 + k_j$ and $q_k = \ell_1+ \ell_2 + k_k$,
are classified into three different sets, $i \in \alpha_1$, $j \in \alpha_2$ and $k \in \alpha_3$, as shown in Fig.~\ref{fig:twoloop}. 
The minus sign in front of $\alpha_2$ indicates that the momenta in $\alpha_2$ are reversed to hold
a momentum flow consistent with $\alpha_1$.
The dual representation in \Eq{eq:LTDtwoloop} spans over the sum of all possible double-cut contributions, with each of the two 
cuts belonging to a different set. At higher orders, the iterative application of LTD introduces a number of
cuts equal to the number of loops~\cite{Bierenbaum:2010cy}. 
The dual amplitudes are thus tree-level like objects to all orders, and can even be related to the forward limit 
of tree-level amplitudes~\cite{Baadsgaard:2015twa,CaronHuot:2010zt}. However, 
neither \Eq{eq:LTDoneloop}, nor \Eq{eq:LTDtwoloop}, or their higher order generalization~\cite{Bierenbaum:2010cy}, 
has been deducted from the forward limit of tree-level amplitudes and they are therefore free of their potential 
spurious singularities. 

A decisive feature that allows us to progress from scattering amplitudes to cross-sections and other physical observables
in the LTD representation is the fact that all IR and physical threshold singularities of the dual amplitudes 
are restricted to a compact region of the loop three-momenta~\cite{Buchta:2014dfa}. This is essential to establish 
a mapping between the real and virtual kinematics in order to locally cancel the IR singularities without the need for subtraction counter-terms, as done
in the Four-Dimensional Unsubtraction (FDU) approach~\cite{Hernandez-Pinto:2015ysa,Sborlini:2016gbr,Sborlini:2016hat}.
This is remarkably the place where the dual $\ii$ prescription plays its main role by encoding efficiently and correctly 
the causal effects of the loop scattering amplitudes. A general analysis of causality~\cite{Buchta:2014dfa,Tomboulis:2017rvd} necessarily requires studying in detail the interplay of the dual prescription among different dual propagators. 

In this letter, we extend explicitly to two loops the analysis of the singular structure of one-loop amplitudes in the LTD 
framework presented in Ref.~\cite{Buchta:2014dfa}, and generalize it to higher orders. 
This allows us to present a comprehensive description of unitarity thresholds and anomalous thresholds 
in the loop-momentum space, and provide generalizable expressions of their singular behavior. 
We show how non-causal unphysical thresholds cancel locally in the forest defined by the 
sum of dual cuts thanks to the momentum dependent $\ii$ prescription of the dual propagators. 
Most importantly, we demonstrate that soft and collinear singularities are always restricted to a compact region 
of the loop three-momentum space. 

\section{Unitarity thresholds, anomalous thresholds and infrared singularities at one loop}

To analyze the singular behavior of one-loop amplitudes in the loop momentum space, 
it is convenient to start by considering the integrand function
\beq
{\cal S}^{(1)}_{ij} = (2\pi \imath)^{-1}\, G_D(q_i;q_j) \,  \td{q_i} + (i \leftrightarrow j)~, 
\eeq
representing the sum of two single-cut dual contributions. 
The singularities of the function ${\cal S}^{(1)}_{ij}$ are encoded through the set of conditions
\beq
\lambda^{\pm\pm}_{ij} = \pm q_{i,0}^{(+)}  \pm q_{j,0}^{(+)}  + k_{ji,0} \to 0~. 
\label{eq:genericoneloop}
\eeq
where $q_{r,0}^{(+)} = \sqrt{\qb_r^2+m_r^2}$, with $r\in\{i,j\}$, are the on-shell loop energies. 
There are indeed only two independent limits that determine the position of the singularities
in the loop momentum space. The limit $\lambda^{++}_{ij}\to 0$ occurs in the intersection of the 
forward on-shell hyperboloid (positive energy mode) of one propagator with the backward on-shell 
hyperboloid (negative energy mode) of the other. The solution to \Eq{eq:genericoneloop} for $\lambda^{++}_{ij}\to 0$ requires
\beq
k_{ji}^2 - (m_j+m_i)^2 \ge 0~, 
\eeq
with $k_{ji,0}<0$. This means that the $q_j$ propagator has to be in the future of the $q_i$ propagator with both propagators causally connected ($\lambda^{--}_{ij}\to 0$ with $k_{ji,0}>0$ represents the complementary solution).
For massless propagators, and light-like separation, $k_{ji}^2=0$, 
the singular surface pinches to a collinear singularity along a finite segment. In both cases, massive or massless, 
\beq
\lim_{\lambda^{++}_{ij}\to 0} {\cal S}^{(1)}_{ij} = 
\frac{\theta(-k_{ji,0}) 
 \theta( k_{ji}^2 - (m_i+m_j)^2)} 
{\, x_{ij} (- \lambda^{++}_{ij}-\ii k_{ji,0})}
+{\cal O}\left((\lambda_{ij}^{++})^0\right)~,
\eeq
with $x_{ij}=4 \, q_{i,0}^{(+)} q_{j,0}^{(+)}$. It should be noted that the limit $\lambda^{++}_{ij}\to 0$ represents the usual unitarity threshold. Since it involves one backward and one forward on-shell hyperboloid, this situation is equivalent to two physical particles propagating in the same direction in time. Moreover, the $\ii$ prescription is exactly the same as in the Feynman representation, and, in particular for IR singularities, the on-shell energy is bounded by the energy of external momenta, $q_{r,0}^{(+)} \le | k_{ji,0} |$ with $r \in \{i,j\}$.

The other potential singularity occurs for $\lambda^{+-}_{ij}\to 0$, with 
\beq
k_{ji}^2 - (m_j-m_i)^2 \le 0~.
\label{eq:forwardforward}
\eeq
It generates unphysical thresholds in each of the dual components of ${\cal S}^{(1)}_{ij}$, but the sum over the two single-cut dual 
contributions is not singular
\beq
\lim_{\lambda^{+-}_{ij}\to 0} {\cal S}^{(1)}_{ij} =
{\cal O}\left((\lambda_{ij}^{+-})^0\right)~.
\label{eq9}
\eeq
The cancellation of this integrand singularity is fully local due to the change of sign in the dual prescription, 
$ q_{j,0}^{(+)}\, G_D(q_i;q_j) |_{\lambda^{+-}_{ij}\to 0} = - q_{i,0}^{(+)}\, G_D(q_j;q_i) |_{\lambda^{+-}_{ij}\to 0} $,
and is not affected by other propagators because 
\beq
\lim_{\lambda^{+-}_{ij} \to 0} G_D(q_j; q_k) = \lim_{\lambda^{+-}_{ij}\to 0} G_D(q_i; q_k)~.
\label{eq10}
\eeq
This second configuration corresponds to the on-shell emission and on-shell reabsorption of one virtual particle. 
The complete local cancellation of this integrand singularity would not occur if all the propagators were Feynman propagators. 
As physics cannot depend on the used representation of the loop amplitude, this mismatch in the $\ii$ prescription 
is compensated in the Feynman Tree Theorem~\cite{Feynman:1963ax} by the multiple-cut contributions.

The LTD representation in \Eq{eq:LTDoneloop} is obtained by defining the momentum flow of the loop anti-clockwise 
and then by closing the Cauchy integration contour of the complex loop energy in the lower half-plane. 
Alternatively, one can close the contour in the upper half-plane and select the modes with negative energy, 
which is equivalent to reversing the momentum flow of the loop. The position of the unitarity thresholds in the loop-momentum space is invariant under this transformation because physics cannot depend on the used dual representation.
However, the position of the unphysical thresholds depends on the choice of the momentum flow. 
Explicitly, in \Eq{eq:forwardforward} we can distinguish between two scenarios. In the space-like case, $k_{ji}^2<0$, unphysical thresholds occur in the intersection of the two forward on-shell hyperboloids and in the intersection 
of the two backward on-shell hyperboloids. Whereas, in the time-like configuration, $0\le k_{ji}^2 \le (m_i-m_j)^2$, unphysical 
thresholds occur only in the intersection of either the forward or the backward on-shell hyperboloids.
For example, the dual representation of the two-point function 
\beq
\int_{\ell_1} \frac{1}{((\ell_1-p)^2-m_1^2+\ii)(\ell_1^2-m_2^2+\ii)}~,
\eeq
with $p=(p_0,{\bf 0})$ and $m_2 < p_0+m_1$ is free of unphysical thresholds, but the integrand with 
inverted momentum flow is not. Of course, the integral is invariant under the transformation $\ell_1 \to -\ell_1$.
This is worth noticing because the proposal of considering both the positive 
and negative modes together in an LTD approach~\cite{Runkel:2019yrs} does not change the physics but inefficiently increases
the number of necessary on-shell cuts to describe a given loop amplitude, and
it also proliferates the number of unphysical singularities of the integrand, in particular at higher orders.

We are now in the position to discuss the anomalous thresholds. 
For internal propagators with real masses, anomalous thresholds at one loop occur when more than two of them go on-shell simultaneously. 
In particular, for three propagators we should analyze the integrand function
\beq
{\cal S}^{(1)}_{ijk} = (2\pi \imath)^{-1}\, G_D(q_i;q_k) \, G_D(q_i;q_j) \,  \td{q_i} + {\rm perm.}~.
\label{eq:anomalousS}
\eeq
The potential singularity of \Eq{eq:anomalousS} arising in the intersection of the three forward on-shell hyperboloids 
cancels again locally among the dual contributions~\cite{Buchta:2014dfa}. In order to generate a physical effect, 
we need to consider the intersection of one forward with two backward on-shell hyperboloids, or two forward with one backward. 

Explicitly, in the double limit $\lambda^{++}_{ij}$ and $\lambda^{++}_{ik} \to 0$ with $k_{ji,0}$ and $k_{ki,0}$ negative,
\bea
\lim_{\lambda^{++}_{ij}, \lambda^{++}_{ik} \to 0} {\cal S}^{(1)}_{ijk}  &=& 
\frac{1} {x_{ijk}}
\prod_{r=j,k} \frac{\theta(-k_{ri,0})
\, \theta( k_{ri}^2 - (m_i+m_r)^2) 
} {(-\lambda^{++}_{ir}-\ii k_{ri,0})} \nn \\
&+&{\cal O}\left((\lambda_{ij}^{++})^{-1}, (\lambda_{ik}^{++})^{-1}\right)~,
\eea
with $x_{ijk}=8 \, q_{i,0}^{(+)} q_{j,0}^{(+)} q_{k,0}^{(+)}$. As expected from the discussion of the cancellation of unphysical thresholds, Eqs. (\ref{eq9}) and (\ref{eq10}), the leading term is free of singularities in $\lambda^{-+}_{jk}$. This is also true for the next terms in the expansion, even though $\lambda^{-+}_{jk} = \lambda^{++}_{ik} - \lambda^{++}_{ij}$. 

Again, the local cancellation of the $\lambda^{-+}_{jk}$ singularity occurs thanks to the momenta dependent 
$\ii$ dual prescription, and the remaining singularities are described by causal $+\ii$ contributions. 
If all the uncut propagators would be Feynman, a mismatch would be generated.
Converserly, the anomalous threshold generated in the intersection of
one backward with two forward on-shell hyperboloids, $\lambda^{++}_{ik}$ and $\lambda^{++}_{jk}\to 0$ with 
$\lambda^{+-}_{ij} = \lambda^{++}_{ik} - \lambda^{++}_{jk}$, is free of singularities in $\lambda^{+-}_{ij}$. 
This configuration also describes a soft singularity in the limiting case of massless partons, with e.g. $q_{i,0}^{(+)} \to 0$. 
Notice that for a soft singularity to be generated the participation of three propagators is necessary due 
to the soft suppression of the integration measure.
The extension of the discussion to anomalous box singularities is straightforward from the results presented here. 

The analysis presented here allows us to move on to the discussion of the two-loop case.

\section{Unitarity thresholds, anomalous thresholds and infrared singularities at two loops}

The characterization of singularities arising at two loops from two or more propagators of the same subset $\alpha_r$ 
is a replica of the one-loop case and does not need further discussion. 
The genuine two-loop IR and threshold singularities arise when the propagator that eventually goes
on shell and the two other on-shell propagators belong to a different subset each. 
In this case, we have to consider the forest defined by all the possible permutations. 
Inspired by the dual representation in~\Eq{eq:LTDtwoloop}, we define 
the following integrand function 
\bea
{\cal S}^{(2)}_{ijk} &=& (2\pi \imath)^{-2}\,  \\ &\times& \left[ G_D(q_j; q_k) \, \td{q_i,q_j} +  G_D(-q_j;q_i) \, \td{-q_j,q_k} \right. \nn \\ 
&+&  \left. \left[  G_D(q_k; q_j) + G_D(q_i;-q_j) - G_F(q_j) \right] \, \td{q_i,q_k} \right]~,  \nn 
\eea
with the shorthand notation $\td{q_r,q_s} = \td{q_r} \td{q_s}$, and 
$i \in \alpha_1$, $j \in \alpha_2$ and $k \in \alpha_3$. The position of the singularities of ${\cal S}^{(2)}_{ijk}$ 
in the loop-momentum space are determined by the set of conditions~\cite{Driencourt-Mangin:2019aix}
\beq
\lambda^{\pm\pm\pm}_{ijk} = \pm q_{i,0}^{(+)}  \pm q_{j,0}^{(+)}  \pm q_{k,0}^{(+)} + k_{k(ij),0} \to 0~,
\label{eq:twoloopconditions}
\eeq
where $k_{k(ij),0} = q_k-q_i-q_j$ depends on external momenta only, 
with our choice of the momentum flow (see Fig.~\ref{fig:twoloop}).
Now, the unitarity threshold is defined by the limit
\bea
\lim_{\lambda^{+++}_{ijk}\to 0} {\cal S}^{(2)}_{ijk} &=& 
\frac{\theta(-k_{k(ij),0})
\, \theta( k_{k(ij)}^2 - (m_i+m_j+m_k)^2)
}{x_{ijk} (- \lambda^{+++}_{ijk}-\ii k_{kj,0})} \nn \\
&+&{\cal O}\left((\lambda_{ijk}^{+++})^0\right)~.
\eea
Notice that $k_{kj,0}$ depends
on the energy of the loop momentum $\ell_1$, however, 
\beq
\left. k_{kj,0} \right|_{\lambda^{+++}_{ijk} \to 0} = q_{i,0}^{(+)} + k_{k(ij),0} < 0~,
\eeq
therefore $-\ii k_{kj,0} = +\ii$ on the physical threshold. We can interpret this singularity as the 
intersection of the backward on-shell hyperboloid of $q_k$ with the forward on-shell hyperboloids 
of $q_i$ and $q_j$. The latter are independent of each other as each of them depends on a different loop momentum. In other words, the three internal physical momenta flow in the same direction in time, which is equivalent to the unitarity cut. Further, this configuration generates a triple collinear singularity for massless partons and light-like separation, $k_{k(ij)}^2=0$. Likewise, the complementary solution with $k_{k(ij),0} > 0$ would generate 
IR or threshold singularities in the limit $\lambda^{---}_{ijk} \to 0$. In any of the two cases discussed so far, the on-shell energies are limited by the energy of the external partons, 
$q_{r,0}^{(+)} \le |k_{k(ij),0}|$ with $r\in\{i,j,k \}$. Since this is an essential condition for the implementation 
of FDU~\cite{Hernandez-Pinto:2015ysa,Sborlini:2016gbr,Sborlini:2016hat} at higher orders this observation constitutes 
one of the main results of this letter. 

Similarly to the one-loop case, there are other potential singularities 
at $\lambda^{++-}_{ijk}\to 0$ and $\lambda^{+--}_{ijk}\to 0$. 
However, those singularities cancel locally in the sum of all the dual components of
${\cal S}^{(2)}_{ijk}$. This is again possible because, even though the dual prescriptions
depend on the loop momenta, the following conditions are fulfilled in each of the cases
\bea
\lambda^{++-}_{ijk}  = 0~, \qquad  \to \qquad q_{k,0}^{(+)} - k_{k(ij),0}  > 0~, \nn \\
\lambda^{+--}_{ijk} = 0~,  \qquad  \to \qquad q_{i,0}^{(+)} + k_{k(ij),0} > 0~,   
\eea
in such a way that all the contributions conspire to match their $\ii$ prescriptions
on the singularity.

Let us stress that the on-shell conditions are determined by a set of equations that are linear in 
the on-shell loop energies to all orders (e.g. \Eq{eq:twoloopconditions}). 
In a recent paper~\cite{Runkel:2019yrs}, an alternative dual prescription has 
been proposed which is quadratic in the loop energies for a subset of the internal propagators at two loops 
(and cubic at three loops, and so on). This new prescription is difficult to reconcile with the linear conditions, while 
within our representation the dual prescriptions are always linear and acquire a well defined sign on the singularities.

\begin{figure}[th]
\begin{center}
\includegraphics[width=0.45\textwidth]{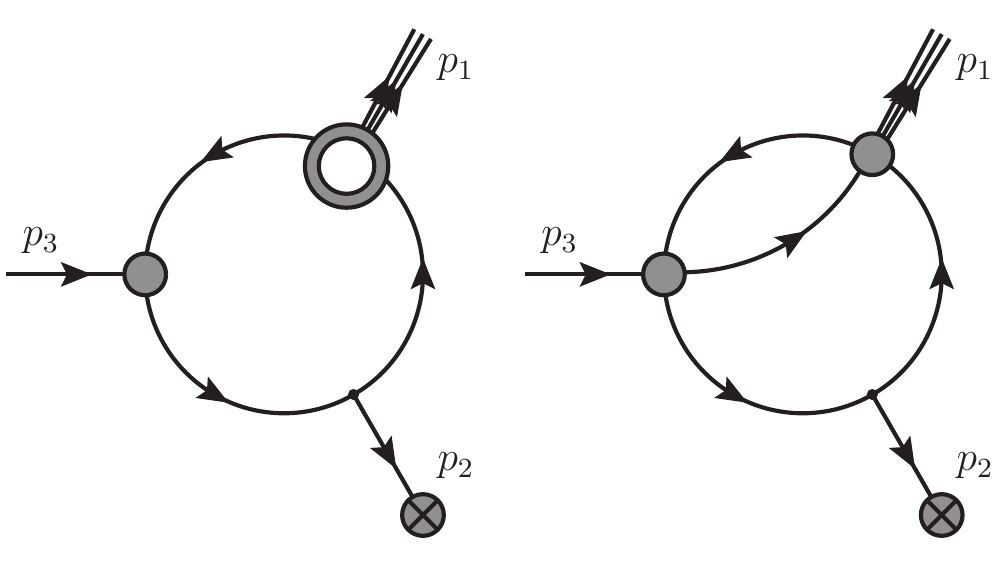}
\caption{Prototype two-loop Feynman diagrams with anomalous thresholds.
\label{fig:anomalous}}
\end{center}
\end{figure}

Concerning anomalous thresholds, we will consider any configuration that involves more on-shell propagators than those appearing in the unitarity cuts. Again, we can distinguish the case where the propagators involved 
belong to the same subset $\alpha_r$, from the genuine two-loop case where it is necessary to 
consider propagators going simultaneously on shell from the three different sets. 
Prototype configurations are illustrated in Fig.~\ref{fig:anomalous}. 
Anomalous thresholds are generated in either of the two benchmarks shown in Fig.~\ref{fig:anomalous}
with the participation of the two propagators adjacent to the external momentum $p_2$, 
with $p_2=q_{i_2}-q_{i_1}$, $\{i_1,i_2\} \in \alpha_1$, satisfying~\Eq{eq:forwardforward}.
Specifically, for the right diagram the anomalous threshold appears when four propagators become singular (i.e. there are two cut ones and two additional internal momenta become on-shell due to kinematics) which we see in the limit
\bea
&& \lim_{\lambda^{+++}_{i_1jk}, \lambda^{+++}_{i_2jk} \to 0} {\cal S}^{(2)}_{i_1i_2jk}  = \nn \\ && 
\frac{1} {x_{i_1 i_2 jk}}
\prod_{i=i_1 i_2} \frac{\theta(-k_{k(ij),0})
 \, \theta(k_{k(ij)}^2-(m_i+m_j+m_k)^2)
} {(-\lambda^{+++}_{ijk}-\ii k_{kj,0})} \nn \\
&& + {\cal O}\left((\lambda_{i_1jk}^{+++})^{-1}, (\lambda_{i_2jk}^{+++})^{-1}\right)~,
\eea
with the obvious generalization of the notation. 
Other anomalous threshold configurations can easily be inferred from these results.
Furthermore, the threshold and singular structure at higher orders can be deducted from 
the explicit dual representations reported in Ref.~\cite{Bierenbaum:2010cy}.

Finally, let us comment that \Eq{eq:LTDtwoloop}, and the equivalent expressions at higher orders,  
can be rewritten in such a way that all the dual prescriptions become independent of the 
loop momenta by introducing additional dual cuts ~\cite{Bierenbaum:2010cy}. However, 
at two-loops and if the set $\alpha_2$ is composed by a single propagator, 
the extra dual cuts can be reabsorbed and the dual 
representation reduces to 
\bea
&&{\cal A}^{(2)}(\{p_n\}_N) = \int_{\ell_1}\int_{\ell_2} {\cal N}(\ell_1,\ell_2,\{p_n\}_N) \otimes \nn \\ 
&& \qquad [ G_D(\alpha_1) \, G_D(\alpha_2) \, G_F(\alpha_3) 
+ G_F(\alpha_1)  \, G_D(-\alpha_2) \, G_D(\alpha_3) \nn \\
&& \qquad + G_D(\alpha_1) \,G_F^*(\alpha_2) \, G_D(\alpha_3) ]~,
\label{eq:LTDalternative}
\eea
with 
\bea
G_F^*(\alpha_2)  = G_F^*(q_j) =\frac{1}{q_j^2-m_j^2-\ii}~. 
\label{eq:gstar}
\eea
As it cannot be otherwise, the threshold and singular solutions to \Eq{eq:LTDalternative} are in full agreement with \Eq{eq:LTDtwoloop}
due to a similar local matching of the dual prescriptions.

In summary, we have presented the first comprehensive description of the singular structure of scattering amplitudes 
in the LTD formalism at higher orders. The LTD representation allows for a detailed discussion directly in the loop 
momentum space. Contrary to the Landau equations~\cite{Landau:1959fi}, 
it does not rely on the Feynman parametrization and keeps track of the imaginary prescription 
of internal propagators in a consistent way. This consistency is needed for the proof that there is a perfect local cancellation 
of non-causal or unphysical thresholds in the forest defined by the sum of all the on-shell dual 
contributions. The remaining causal thresholds, and in particular the soft and collinear singularities, 
which are described as the limiting case of threshold singularities, are restricted to a compact region 
of the loop three-momenta space. This feature is essential to prove that it is possible to establish a 
mapping between the kinematics of the virtual and real contributions to physical observables in such 
a way that the cancellation of IR singularities occurs locally, as defined in the FDU approach.

{\it Acknowledgements:} We thank David Broadhurst for pointing us about the open question of anomalous thresholds at higher orders.
This work is supported by the Spanish Government  (Agencia Estatal de Investigaci\'on) and ERDF funds from European
Commission (Grants No. FPA2017-84445-P and SEV-2014-0398), Generalitat Valenciana (Grant No. PROMETEO/2017/053),
Consejo Superior de Investigaciones Cient\'{\i}ficas (Grant No. PIE-201750E021) and the COST Action CA16201 PARTICLEFACE. 
JP acknowledges support from "la Caixa" Foundation (ID 100010434, LCF/BQ/IN17/11620037), and from the European Union's H2020-MSCA 
Grant Agreement No. 713673, 
NSR from CONACYT, 
JJA from Generalitat Valenciana (GRISOLIAP/2018/101)
and WJT from the Spanish Government (FJCI-2017-32128). 

\bibliography{refs}
\end{document}